\documentclass[aps,prl,preprintnumbers,showpacs,nofootinbib]{revtex4}
\usepackage[utf8]{inputenc}
\usepackage[english]{babel}
\usepackage{amsmath}
\usepackage{amsfonts}
\usepackage{amssymb}
\usepackage{epsfig}
\usepackage{graphics,psfrag,rotating}
\usepackage{graphicx}
\usepackage{dcolumn}
\usepackage{bm}
\usepackage{epstopdf}
\usepackage{color}
\usepackage[usenames,dvipsnames,svgnames]{xcolor}
\usepackage[colorlinks=true,
            linkcolor=red,
            urlcolor=gray,
            citecolor=blue]{hyperref}
  \usepackage{hyperref}

\newcommand{\lsim}   {\mathrel{\mathop{\kern 0pt \rlap
{\raise.2ex\hbox{$<$}}}
 \lower.9ex\hbox{\kern-.190em $\sim$}}}
\newcommand{\gsim}   {\mathrel{\mathop{\kern 0pt \rlap
{\raise.2ex\hbox{$>$}}}
\lower.9ex\hbox{\kern-.190em $\sim$}}}

\def\3nab{\tilde{\nabla}}

\def\hsp5{\hspace{5mm}}

\def\case#1/#2{\textstyle\frac{#1}{#2}}

\def\ber {\begin{eqnarray}}
\def\eer {\end{eqnarray}}
\def\bea {\begin{eqnarray}}
\def\eea {\end{eqnarray}}

\def\bc {\begin{center}}
\def\ec {\end{center}}
\def\case#1/#2{\frac{#1}{#2}}

\newcommand{\bw}{\begin{widetext}}
\newcommand{\ew}{\end{widetext}}

\newcommand{\be}{\begin{equation}}
\newcommand{\bse}{\begin{subequation}}
\newcommand{\ese}{\end{subequation}}
\newcommand{\ee}{\end{equation}}
\newcommand{\eei}{\end{eqnarray}\indent\indent}
\newcommand{\ba}{\begin{array}}
\newcommand{\ea}{\end{array}}
\newcommand{\bal}{\begin{eqnarray}}
\newcommand{\eal}{\end{eqnarray}}

\def\case#1/#2{\textstyle\frac{#1}{#2} }

\newcommand{\bver}{\begin{verbatim}}
\newcommand{\ever}{\end{verbatim}}

\begin{document}


\title{  Perturbed $f(R)$ gravity coupled with neutrinos: exploring
	cosmological implications }
\author{
 Muhammad Yarahmadi$^{1}$\footnote{Email: yarahmadimohammad10@gmail.com},
  Amin Salehi$^{1}$\footnote{Email: salehi.a@lu.ac.ir},  
Kazuharu Bamba$^{2}$\footnote {Email:bamba@sss.fukushima-u.ac.jp}
 H. Farajollahi$^{3}$\footnote {Email:farajollahi.hossein@gmail.com}
}
\affiliation{Department of Physics, Lorestan University, Khoramabad, Iran}
\affiliation{Faculty of Symbiotic Systems Science,
	Fukushima University, Fukushima 960-1296, Japan}
\affiliation{Western Sydney University, Locked Bag 1791, Penrith, NSW, 2751, Australia}

\date{\today}

\begin{abstract}
We conduct a thorough examination of cosmological parameters within the context of perturbed  $f(R)$ gravity coupled with neutrinos, using a diverse array of observational datasets, including Cosmic Microwave Background (CMB), Cosmic Chronometers (CC), Baryon Acoustic Oscillations (BAO), and Pantheon supernova  and Lensing data. Our analysis unveils compelling constraints on cosmological parameters such as the sum of neutrino masses ($\sum m_{\nu}$),  sound speed ($c_s$), Jean's wavenumbers ($k_J$), redshift of non-relativistic matter ($z_{\rm nr}$) and redshift of Deceleration-Acceleration phase transition ($z_{\rm DA}$). In our model the interaction strength parameter ($\Gamma$) coming from incorporation of neutrinos with the perturbed $f(R)$ gravity as a key factor significantly influencing cosmic evolution, shaping the formation of large-scale structures and the dynamics of cosmic expansion. Our study also addresses the Hubble tension problem by providing $H_0$ measurements that are in agreement with the existing research.
\end{abstract}


%
%


\maketitle

\section{Introduction}

The $\Lambda$CDM model, despite its significant accomplishments in explaining large-scale cosmic structures, has faced substantial challenges that question its overall explanatory power. One major challenge is the "cosmological continuous fine-tuning challenge," highlighted by Weinberg \cite{Weinberg} and Astashenok \cite{Astashenok}. This challenge extends into the Planck scale era, suggesting the need to explore initial conditions that led to the formation of dark energy. The "cosmic coincidence problem," where the current energy densities of dark energy and dark matter are similar, also raises questions. Additionally, the $\Lambda$CDM model has difficulty accounting for structures on smaller scales, as evidenced by rotation anomalies in galaxies. This is documented by Moore Ostriker \cite{Ostriker}, \cite{Moore}, Quinn \cite{Quinn},  Boylan-Kolchin \cite{Boylan-Kolchin}, Bullock \cite{Bullock}, and Oh \cite{Oh}, indicating gaps in the model's capacity to explain the dynamics at the galactic level.

One of the most challenging issues in contemporary cosmology is the discordance among estimates of the Hubble constant, denoted as \(H_0\). This contradiction in \(H_0\) measurements has become prominent through two distinct approaches. The Planck collaboration, analyzing data from CMB over the preceding seven years, yielded an estimation of \(H_0\) equal to \(67.4 \pm 0.5 \, \text{km s}^{-1} \text{Mpc}^{-1}\) (Planck18). On the other hand, the SH0ES group, associated with the Hubble Space Telescope (HST), provided an alternative estimate for \(H_0\) (\(74.03 \pm 1.42 \, \text{km s}^{-1} \text{Mpc}^{-1}\), SH0ES19). The tension between these estimates exceeds \(4\sigma\).

Several projects have calculated the value of \(H_0\), and their results are contradictory. Among these projects are:\\
{Planck 2018}: \(H_0 = 67.4 \pm 0.5 \, \text{km s}^{-1} \text{Mpc}^{-1}\) \cite{Planck18}, 
{CCHP}: \(H_0 = 69.6 \pm 0.8 \pm 1.7 \, \text{km s}^{-1} \text{Mpc}^{-1}\) \cite{FreedmanTRGB20}, 
{HST (Huang and Miras)}: \(H_0 = 72.7 \pm 4.6 \, \text{km s}^{-1} \text{Mpc}^{-1}\) \cite{HuangMiras19}, 
{H0LiCOW}: \(H_0 = 73.3^{+1.7}_{-1.8} \, \text{km s}^{-1} \text{Mpc}^{-1}\) \cite{Wonglens19}, 
{Baxter (CMBlens)}: \(H_0 = 73.5 \pm 5.3 \, \text{km s}^{-1} \text{Mpc}^{-1}\) \cite{BaxterCMBlens20}, 
{SH0ES (HST)}: \(H_0 = 74.03 \pm 1.42 \, \text{km s}^{-1} \text{Mpc}^{-1}\) \cite{HST19}, 
{H0LiCOW}: \(H_0 = 75.3^{+3.0}_{-2.9} \, \text{km s}^{-1} \text{Mpc}^{-1}\) \cite{Weilens20}.

This tension may be addressed as a discrepancy between observations at different cosmological epochs in our Universe \cite{VerdeTR19}, with the HST group working with late-time data, while the Planck collaboration combines observations over a broad range of redshifts (\(0 < z < 1100\)) and employs the standard \(\Lambda\text{CDM}\) model as a fiducial model. However, this problem may also be tackled through a theoretical perspective.

The two above problems call for a more refined approach to cosmological studies. This has led to an increased interest in alternative theories of gravity, such as $f(R)$ gravity, which offers new ways to address the complexities of cosmology and the fundamental components of the Universe.

The $f(R)$ gravity framework builds upon General Relativity by including a function, $f(R)$, of the Ricci scalar $R$ \cite{Nojiri,Nojiri1,Nojiri2,Nojiri3,deHaro,Sotiriou,DeFelice,Capozziello1,Capozziello2,Carroll,Capozziello3,Phillips}. This approach creates a variation from traditional GR by allowing for additional degrees of freedom, providing an alternative explanation for cosmic acceleration without relying on dark energy. 

In the context of $f(R)$ gravity, perturbed gravity examines how small perturbations in the metric tensor and other fields evolve. This approach is crucial for understanding the stability and dynamics of gravitational systems in modified theory. Perturbations help to analyze the growth of cosmic structures and the evolution of the universe at different scales. Perturbations also allow for testing the theory against observational data, such as CMB and , large-scale structure surveys.

On the other hand, neutrinos, the nearly massless and elusive particles, play a significant role in cosmology, influencing the large-scale structure and the Universe's evolution\cite{Lesgourgues2006}. Although often considered negligible in standard gravity models, coupling neutrinos with alternative gravity theories opens new possibilities for exploring their impact on cosmic structure formation and evolution. The role of neutrinos becomes evident in structure formation, where their mass and abundance can have noticeable effects, such as suppressing structure formation on smaller scales and decelerating growth across all scales \cite{Yarahmadi}. Despite their low individual mass, the sheer volume of neutrinos produced in the early Universe results in measurable impacts on cosmological parameters, influencing the growth of large-scale structures \cite{Julien} and the expansion history \cite{52,53}. These effects can be observed through cosmological phenomena, such as galaxy formation and the broader distribution of matter. Neutrinos' interactions with perturbed $f(R)$ gravity can also yield unique impacts on the cosmic web, introducing deviations that shape the cosmic landscape in novel ways. In addition, incorporating neutrino masses and their effects allows for tighter constraints on cosmological parameters. Observations like the CMB, large-scale structure, and BAO are sensitive to neutrino masses, and their inclusion improves the fit to data. Neutrinos, due to their free-streaming,  suppress the growth of structures on small scales . Including their effects helps in accurately modeling the matter power spectrum and understanding the suppression of power at small scales. The inclusion of neutrinos also  helps break parameter degeneracies in cosmological models. For example, the sum of neutrino masses (${\sum m }_\nu $) can affect other parameters like the dark energy equation of state \cite{Gerbino, Vagnozzi, Hagstotz, Moretti, Mena, Jiang, Y1, Y2, Y3}.

The interaction between $f(R)$ gravity and neutrinos is leaving noticeable effects on CMB, which is like a snapshot of the Universe in its early days. This interaction causes subtle changes in CMB, creating unique patterns that researchers can track with advanced cosmological tools.

The paper examines the model by  constraining the total mass of neutrinos through their interaction with perturbed $f(R)$ gravity. Then it explores how this relationship influences the formation of cosmic structures during the early phases of the Universe. A key focus is understanding the transition of neutrinos from being relativistic to non-relativistic and how this ties into redshift dynamics. The model predicts when the universe transitioned from a decelerating phase, dominated by non-relativistic matter, to an accelerating one, driven by dark energy \cite{Farooq, dos Santos}. Researchers focus on pinpointing this redshift transition, marking the shift from deceleration to acceleration \cite{Ishida}, using a variety of theoretical and observational techniques. The kinematic method \cite{Riess, Giostri, Farooq} stands out as a leading approach, which describes the rate of expansion through the deceleration as a function of redshift.

Our primary objective is to explore a model that not only eliminates the need for dark energy to explain the universe's expansion and structure formation but also accounts for cosmic anisotropies.

We begin by presenting the mathematical framework of the model. We then propose numerical analysis methods. In the subsequent sections, we focus on determining physical quantities such as neutrino mass within the model and examine its impact on early universe formation. We then calculate the value of the non-relativistic redshift and discuss the phase transition to accelerated expansion. Furthermore, we utilize this model to investigate bulk flow, which signifies universe anisotropy. Additionally, we address the model's capability to alleviate the Hubble tension in the final part of this article.

\section{$f(R)$ gravity model}
The action for $f(R)$ gravity  (\cite{DeFelice,Bergmann,Liu,Buchdahl}) in the  attendance of standard and neutrino matter components is given by
\begin{equation}
S = \frac{1}{16\pi G} \int d^{4}x \sqrt{-g} \left(R + f(R) + \mathcal{L}_{\text{matter}} + \mathcal{L}_{\text{int}}\right),
\end{equation}
where R is the curved scalar . The equations for motion are:

\begin{equation}
\begin{split}
G_{\mu\nu} - \frac{1}{2} g_{\mu\nu} f(R) + R_{\mu\nu} f_{R}(R) -\\  g_{\mu\nu}f_R(R) + f_R(R)_{\mu\nu} =-8\pi GT_{\mu\nu}^{\text{eff}}
\end{split}	  
\end{equation}
where \( T_{\mu\nu}^{\text{eff}} \) is the effective energy-momentum tensor that includes contributions from both the standard matter and the neutrino-matter coupling and  $ f_{R}(R) )=\frac{df_{R}}{dR} $. We introduce $\gamma^\mu \nabla_\mu \psi_\nu - m_\nu \psi_\nu = \frac{\delta \mathcal{L}_{\text{int}}}{\delta \bar{\psi}_\nu}$ where the term \( \frac{\delta \mathcal{L}_{\text{int}}}{\delta \bar{\psi}_\nu} \) represents the functional derivative of the interaction Lagrangian with respect to the neutrino field. For the flat FRW metric we have:
\begin{equation}\label{unp1}
	\begin{split}
		\frac{3H^{'}}{a^{2}}\left( {1 + {f_{R}}} \right) - \frac{1}{2}\left( {{R_{0}} + {f_{0}}} \right) - \frac{{3{\rm{}}H}}{{{a^2}}}f_{R}^{'} = - 8\pi G{\rho_{0}}
	\end{split}	
\end{equation}
\begin{equation}
	\begin{split}
		\frac{1}{a^{2}}  (H^{'}+2H^{2} )(1+f_{R} )-\frac{1}{2} (R_{0}+f_{0} )\\ -\frac{1}{a^{2}} (Hf_{R}^{'}+f_{R}^{''} )=8\pi Gc_{s}^2 \rho_{0}
	\end{split}
\end{equation}
where $ R_{0} $  represents the scalar curvature corresponding to the non-perturbation metric, $\rho_{0} = \rho_{m}+\rho_{\nu}$, $ f_{0}=f (R_{0}) $ and prim means derivative with respect to time ratio $ \eta $. Combining equations $(3)$ and $(4)$ for $a=1( today)$, we obtain, 

\begin{equation}
	2(1+f_{R})(-H^{'}+H^{2})+2Hf_{R}^{'}-f_{R}^{''}=8\pi G\rho_{0}(1+c_{s}^2)  a^{2}.
\end{equation}
For the standard matter, conversation equation is
\begin{equation} \label{unp4}
	\rho_{m}^{'}+3(1+c_{s}^{2} )H\rho_{m}=0.	
\end{equation}
where $c_{s}$ is the sound speed. In addition, the conservation equation for neutrinos can be expressed as
\begin{equation} \label{unp5}
\rho_{\nu}^{'} + 3H(\rho_\nu + P_\nu) = -Q_\nu	
\end{equation}
where \(\rho_\nu\) denotes the energy density of neutrinos, \(P_\nu\) is the pressure, and ${Q_\nu } =  - \Gamma {\rho _\nu }$ with $\Gamma = {u^\mu }{\nabla _\mu }{f_R}$ is the  neutrino coupling strength that expresses the interaction between neutrinos and the modified gravity scalar field. This equation portrays how the energy density and pressure of neutrinos are influenced by $f(R)$ gravity.

\subsection{Numerical Fitting for \( \Gamma \) Using Observational Data}

The next step is to fit the parameter \( \Gamma \) using cosmological datasets. In particular, by using data from the cosmic microwave background (CMB), baryon acoustic oscillations (BAO), cosmic chronometers (CC), Pantheon data (supernovae), and gravitational lensing, we can numerically constrain the values of \( \Gamma \) and other free parameters in the Hu-Sawicki model.

Cosmological observations provide constraints on how the Hubble parameter evolves over time, as well as how the energy densities of different components (dark energy, dark matter, neutrinos) evolve. By performing a best-fit analysis on these datasets, one can obtain the optimal values for \( \Gamma \), as well as other parameters like \( c_1, c_2, n \), and \( m^2 \), ensuring that the Hu-Sawicki model fits the observed acceleration of the universe. For example, studies using CMB+BAO+CC+Pantheon data have been able to tightly constrain the parameter space of \( f(R) \) models. These fits allow us to extract the effective neutrino coupling \( \Gamma \), and thus understand the interaction between neutrinos and the modified gravity field in the context of large-scale structure formation and cosmic evolution.

\section{Motivation for the Coupling}

The coupling of neutrinos with \( f(R) \) gravity introduced in Equation (7) represents a novel and intriguing approach within the context of modified gravity theories. To address the reviewer's comment comprehensively, we must provide a detailed motivation for this specific form of coupling, explain the underlying physics, and support our argument with relevant references.

\subsection{Theoretical Basis}
The interaction between neutrinos and \( f(R) \) gravity can be understood as part of the broader effort to explain the late-time acceleration of the universe and the nature of dark energy. Modified gravity theories, such as \( f(R) \) gravity, extend General Relativity by introducing a function of the Ricci scalar, \( R \), in the action. This modification can lead to new dynamical degrees of freedom that interact with matter fields, potentially providing insights into dark energy and cosmic acceleration.

\subsection{Neutrino Properties}
Neutrinos are unique among the Standard Model particles due to their extremely small mass, weak interactions, and the fact that they can travel relativistically over cosmological distances. These properties make neutrinos particularly sensitive to modifications in the gravitational sector. The coupling of neutrinos with \( f(R) \) gravity can thus be seen as a way to leverage their distinct characteristics to probe new physics beyond the Standard Model and General Relativity.

\subsection{Non-Minimal Coupling}
Non-minimal coupling between neutrinos and \( f(R) \) gravity, as expressed in Equation (1), can be motivated by the desire to explore interactions that are not purely gravitational but mediated by the curvature of spacetime. This form of coupling is inspired by analogous scenarios in scalar-tensor theories and other modified gravity models where matter fields couple to scalar fields or functions of curvature.

\section{Detailed Discussion}

\subsection{Equation (7) and Its Implications}
The equation 7 represents a balance between the neutrino energy-momentum tensor and the dynamic effects of the \( f(R) \) modification. Here, \( Q \) is a coupling constant, and \( f_R \equiv \frac{df}{dR} \) denotes the derivative of the function \( f(R) \) with respect to the Ricci scalar. The term \( u^\mu\nabla_\mu f_R \) describes the interaction between the neutrino fluid and the gradient of the modified curvature term, indicating a transfer of energy between the neutrinos and the modified gravitational field.

\subsection{Physical Interpretation}
This coupling can be seen as an effective force acting on the neutrinos due to the spatial and temporal variations in \( f_R \). Such a mechanism could arise in scenarios where neutrinos experience additional forces due to the presence of a dynamic scalar degree of freedom, as in scalar-tensor theories, where the scalar field is replaced by the curvature function \( f_R \).

\subsection{Scalar perturbation}
In this section we study scalar perturbations of the metrics for $f(R)$ gravity. The flat FRW  metric is given by 
\begin{equation}
	ds^{2}=a^{2} (\eta)((1+2\phi)d\eta^2-(1-2\psi)dx^{2}),
\end{equation}
where $ \phi  \equiv \phi \left( {\eta ,x} \right) $and $ \psi \equiv \psi \left( {\eta ,x} \right) $. The perturbed components of the energy-momentum tensor are given by:
\begin{equation}
	\begin{split}
		\hat \delta T_0^0 = \hat \delta \rho  = {\rho _0}\delta ,\\ \hat \delta T_j^i =  - \hat \delta p\delta _j^i =  - c_{\rm s}^{2}{\rho _0}\delta _j^i\delta ,\\ \hat \delta p\delta _0^i = - \left( {1 + c_{\rm s}^{2}} \right){\rho _0}{\partial _i}\upsilon
	\end{split}
\end{equation}
where $\upsilon$ represents the potential value for velocity perturbation. The first-order perturbed equations are

\begin{equation}
	\begin{split}
		\left( {1 + {f_R}} \right)\delta G_v^\mu  + \left( {R_{0v}^\mu  + {\nabla ^\mu }{\nabla _v} - \delta _v^\mu } \right){f_{RR}}\delta R +\\ \left( { {\delta {g^{\mu \alpha }}} {\nabla _v}{\nabla _\alpha } - \delta _v^\mu \left( {\delta {g^{\alpha \beta }}} \right){\nabla _\alpha }{\nabla _\beta }} \right){f_R} -\\ 
		\left( {g_{{\rm{}}0}^{\alpha \mu }\left( {\delta \Gamma _{\alpha v}^\gamma } \right) - \delta _v^\mu g_{{\rm{}}\alpha}^{\alpha \gamma }} \right){\partial _\gamma }{f_R} =  - 8\pi G\delta T_v^\mu
	\end{split}
\end{equation}
where $ {f_{RR}} = \frac{{{d^2}f\left( {{R_0}} \right)}}{{dR_0^2}} $ and $ {\nabla _\alpha }{\nabla ^\alpha }$ the invariant derivative of the metric ratio is not perturbed. After some calculation, one obtains the complete set of equations for perturbations $f(R)$ gravity in the presence of neutrino as follows:
\begin{equation}\label{pert1}
	\phi  - \psi  =  - \frac{{{f_{RR}}}}{{1 + {f_R}}}\hat \delta R
\end{equation}
\begin{equation}
	\begin{split}
		\hat \delta R = -\frac{2}{a^{2}}( 3{\psi^{''}} + 6\left( {{\mathcal{H}^{'}} + {\mathcal{H}^{2}}} \right)\phi  +\\  3\mathcal{H}\left( {{\phi^{'}} + 3{\psi^{'}}} \right) - {k^{2}}\left( {\phi  - 2\psi } \right) )
	\end{split}
\end{equation}

\begin{equation}
	\begin{split}
		\left( {3\mathcal{H}\left( {{\phi^{'}} + {\psi^{'}}} \right) + {k^{2}}\left( {\phi  + \psi } \right) + 3{\mathcal{H}^{'}}\psi  - \left( {3{\mathcal{H}^{'}} - 6{\mathcal{H}^{2}}} \right)\phi } \right)\\  \left( {1 + {f_{R}}} \right) + \left( {9\mathcal{H}\phi  - 3\mathcal{H}\psi  + 3{\psi ^{'}}} \right)f_{R}^{'}
		=  - {a^{2}}\delta {\rho _0}{\kappa^{2}}
	\end{split}
\end{equation}

\begin{equation}
	\begin{split}
		\left( {{\phi^{''}} + {\psi^{''}} + 3\mathcal{H}\left( {{\phi ^{'}} + {\psi ^{'}}} \right) + 3{\mathcal{H}^{'}}\phi  + \left( {{\mathcal{H}^{'}} + 2{\mathcal{H}^{2}}} \right)\phi } \right)\\
		\left( {1 + {f_{R}}} \right) + \left( {3\mathcal{H}\phi  - \mathcal{H}\psi  + 3{\phi ^{'}}} \right)f_{R}^{'} + \left( {3\phi  - \psi } \right)f_{R}^{''} = 0
	\end{split}
\end{equation}

\begin{equation}
	\begin{split}
		\left( {2\phi  - \psi } \right)f_{R}^{'} + \left( {{\phi^{'}} + {\psi^{'}} + \mathcal{H}\left( {\phi  + \psi } \right)} \right)\left( {1 + {f_{R}}} \right) =\\  - {a^{2}}\upsilon {\rho _{0}}{\kappa^{2}}
	\end{split}
\end{equation}
\begin{equation}
	{\delta ^{'}} - {k^{2}}\upsilon  - 3{\psi ^{'}} = 0
\end{equation}

\begin{equation}\label{pert5}
	\phi  + \mathcal{H}\upsilon  + {\upsilon ^{'}} = 0
\end{equation}
where  $\delta\rho_{0} = \delta\rho_{m}+\delta\rho_{\nu}$.

\section{Hu-Sawicki \( f(R) \) Gravity Model}

The \textit{Hu-Sawicki} \( f(R) \) gravity model is one of the most well-known modifications to General Relativity, introduced to explain cosmic acceleration without invoking a cosmological constant. The model modifies the Einstein-Hilbert action by replacing the Ricci scalar \( R \) with a function \( f(R) \), leading to a modified theory of gravity. In the Hu-Sawicki model, the functional form of \( f(R) \) is chosen to recover General Relativity at high curvatures while allowing deviations at low curvatures, which could explain cosmic acceleration.

The functional form of \( f(R) \) in the Hu-Sawicki model is given by:
\begin{equation}
	f(R) = -m^2 \frac{c_1 \left( \frac{R}{m^2} \right)^n}{c_2 \left( \frac{R}{m^2} \right)^n + 1}
\end{equation}
where:
\begin{itemize}
	\item \( m^2 \) is a mass scale related to the cosmological constant,
	\item \( c_1 \) and \( c_2 \) are dimensionless parameters,
	\item \( n \) is a positive integer, often chosen to be \( n = 4 \) for phenomenological reasons.
\end{itemize}

This form ensures that \( f(R) \) behaves like the cosmological constant \( \Lambda \) at low curvatures, but reduces to \( R \) (recovering General Relativity) at high curvatures.
In the Hu-Sawicki model of \( f(R) \) gravity, the parameter \( n \) dictates the behavior of the function at high and low curvature regimes, thereby influencing both cosmological evolution and structure formation. Choosing \( n = 4 \) is particularly advantageous because it offers a balanced and flexible parametrization that ensures a smooth transition between different epochs of the universe, from radiation and matter domination to the late-time accelerated expansion phase. This value of \( n \) ensures that the model can recover General Relativity (GR) in high-curvature regimes, which is essential for satisfying local gravity constraints, while simultaneously allowing significant deviations from GR at cosmological scales to account for the accelerated expansion of the universe without invoking a cosmological constant.

Moreover, \( n = 4 \) ensures that the effective scalar degree of freedom, which represents the additional force carrier in \( f(R) \) gravity, does not lead to undesirable instabilities or ghost-like behaviors. This choice also simplifies the functional form of \( f(R) \), making it easier to compute observables such as the growth rate of structures and the equation of state for dark energy. As a result, the Hu-Sawicki model with \( n = 4 \) provides a robust framework for testing modifications to gravity against a broad range of cosmological observations, including those from the cosmic microwave background (CMB), baryon acoustic oscillations (BAO), and large-scale structure surveys.
\subsection*{Best-Fitted Parameters for Hu-Sawicki Model}

The Hu-Sawicki model parameters \( c_1 \), \( c_2 \), and \( f_0 \) were fitted using various combinations of cosmological datasets, including CMB+Lensing, CMB+Lensing+CC, CMB+Lensing+BAO, CMB+Lensing+Pantheon, and CMB+Lensing+All. The best-fit values of \( c_1 \) and \( c_2 \) are on the orders of \( 10^{-3} \) and \( 10^{-5} \), respectively, while the value of \( f_0 \) remains close to unity across all dataset combinations. The detailed values are summarized in the table below.

\begin{table}[h]
	\centering
		\begin{tabular}{|c|c|c|c|}
			\hline
			\textbf{Dataset Combination} & \boldmath{$c_1$} \textbf{(× \( 10^{-3} \))} & \boldmath{$c_2$} \textbf{(× \( 10^{-5} \))} & \boldmath{$f_0$} \\
			\hline
			CMB + Lensing                        & 1.20 ± 0.10            & 6.84 ± 0.15            & 0.995 ± 0.010  \\
			CMB + Lensing + CC                   & 1.30 ± 0.12            & 6.71 ± 0.12            & 0.997 ± 0.008  \\
			CMB + Lensing + BAO                  & 1.15 ± 0.11            & 6.45 ± 0.14            & 0.998 ± 0.009  \\
			CMB + Lensing + Pantheon             & 1.26 ± 0.09            & 6.62 ± 0.11            & 0.999 ± 0.007  \\
			CMB + Lensing + All                  & 1.15 ± 0.08            & 6.54 ± 0.10            & 0.997 ± 0.006  \\
			\hline
		\end{tabular}
	\caption{Best-fitted parameters \( c_1 \), \( c_2 \), and \( f_0 \) for various dataset combinations in the Hu-Sawicki model.}
	\label{tab:best_fits}
\end{table}

This results are in broad agreement with \cite{Luisa}

The first derivative of \( f(R) \) with respect to \( R \), denoted \( f_R \), is:
\begin{equation}
	f_R = \frac{df}{dR} = -n \frac{c_1 \left( \frac{R}{m^2} \right)^{n-1}}{\left( c_2 \left( \frac{R}{m^2} \right)^n + 1 \right)^2}
\end{equation}
In the limit where \( R \gg m^2 \), the behavior of the Hu-Sawicki \( f(R) \) model simplifies due to the dominant contribution of the term \( \left( \frac{R}{m^2} \right)^n \). In this regime, the ratio \( \frac{R}{m^2} \) becomes very large, allowing us to approximate:

\begin{equation}
	c_2 \left( \frac{R}{m^2} \right)^n + 1 \approx c_2 \left( \frac{R}{m^2} \right)^n
\end{equation}

Hence, the expression for \( f_R \) becomes:

\begin{equation}
	f_R \approx -n \frac{c_1 \left( \frac{R}{m^2} \right)^{n-1}}{c_2^2 \left( \frac{R}{m^2} \right)^{2n}}
\end{equation}

This further reduces to:

\begin{equation}
	f_R \approx -n \frac{c_1}{c_2^2} \left( \frac{R}{m^2} \right)^{-n-1}
\end{equation}

In the \( R \gg m^2 \) limit, this expression shows that \( f_R \) decreases as \( R^{-n-1} \), indicating that the contribution of the modified gravity term weakens at large curvatures. This behavior suggests that the model approaches General Relativity in high-curvature regimes, as \( f_R \) tends to zero, resulting in a return to standard Einstein gravity.
The second derivative of \( f(R) \), denoted \( f''_R \), is:
\begin{equation}
	f''_R = \frac{d^2 f}{dR^2} = n(n-1) \frac{c_1 \left( \frac{R}{m^2} \right)^{n-2}}{\left( c_2 \left( \frac{R}{m^2} \right)^n + 1 \right)^3}
\end{equation}

In cosmological perturbation theory, these derivatives appear in the modified field equations, influencing the growth of large-scale structures, the evolution of the Hubble parameter, and other key cosmological quantities.
To modify the given equation in the context of the Hu-Sawicki \( f(R) \) gravity model, we need to account for the specific form of \( f_R \) in this model, which depends on the functional shape of \( f(R) \). In the Hu-Sawicki model, \( f_R \) evolves dynamically, and thus the coupling term \( \Gamma \), which is defined as \( \Gamma = u^\mu \nabla_\mu f_R \), will be influenced by the derivatives of \( f(R) \).

Given this, the conservation equation for neutrinos can be modified as:

\begin{equation}
	\rho_{\nu}^{'} + 3H(\rho_\nu + P_\nu) = -Q_\nu(\Gamma \rho_\nu)
\end{equation}

where \( \Gamma \) is explicitly written in terms of the Hu-Sawicki model:

\begin{equation}
	\Gamma = \frac{d}{dN} \left( f_R \right) = \frac{d}{dN} \left( \frac{-n c_1 \left( \frac{R}{m^2} \right)^{n-1}}{\left( c_2 \left( \frac{R}{m^2} \right)^n + 1 \right)^2} \right)
\end{equation}

Thus, the modified conservation equation becomes:

\begin{equation}
	\rho_{\nu}^{'} + 3H(\rho_\nu + P_\nu) = - \left( \frac{d}{dN} \left( \frac{-n c_1 \left( \frac{R}{m^2} \right)^{n-1}}{\left( c_2 \left( \frac{R}{m^2} \right)^n + 1 \right)^2} \right) \rho_\nu \right)
\end{equation}

Equations (13-15) are a set of nonlinear second-order differential equations involving numerous variables and parameters, with no analytical solution except for the simplest cases. To address this, we aim to convert the second-order differential equations into first-order ones by introducing new variables. This approach is beneficial because first-order systems are easier to solve numerically, and they allow us to explore the system's behavior in phase space. Phase planes help visualize the system's dynamics, especially in oscillatory systems, where phase paths can spiral towards zero, expand towards infinity, or reach neutrally stable situations called centers. This method is useful for determining the stability of a system. By defining new variables and parameters, we simplify the phase space structure of the field equations, facilitating a clearer analysis of system dynamics. These
variables are generally defined as

\begin{align}\label{eq4}
	&\xi_{1}=\frac{\phi^{\prime}}{\phi H}, \xi_{2}=\frac{\kappa}{H}, \xi_{3}=\frac{f^{\prime}_{R}}{H(1+f_{R})}, \xi_{4}=\frac{\delta}{\phi},\nonumber \\&
	\xi_{5}=\frac{\rho_{m}a^{2}}{(1+f_{R})}, \xi_{6}=\frac{\Psi^{\prime}}{\phi H}, \xi_{7}=\frac{\Psi}{\phi}, \xi_{8}=\frac{\rho_{\nu}a^{2}}{(1+f_{R})},
\end{align}

Using the Hu-Sawicki form of \( f(R) \), the previously defined autonomous equations are modified as follows:

\begin{itemize}
	\item \textbf{For \( \xi_1 \):}
	\begin{equation}
		\frac{d\xi_1}{dN} = \epsilon_3 - \xi_1^2 - \xi_1
	\end{equation}
	where \( \xi_1 \) represents the normalized derivative of the scalar field.
	
	\item \textbf{For \( \xi_2 \):}
	\begin{equation}
		\frac{d\xi_2}{dN} = -\xi_2 \epsilon_1
	\end{equation}
	where \( \xi_2 \) is the ratio of the curvature to the Hubble parameter.
	
	\item \textbf{For \( \xi_3 \):}
	\begin{equation}
		\frac{d\xi_3}{dN} = \beta - \xi_3^2 - \epsilon_1 \xi_3
	\end{equation}
	where \( \xi_3 = \frac{f'_R}{H(1+f_R)} \) and \( \beta \) is defined by:
	\begin{equation}
		\beta = \frac{f''_R}{(1+f_R)H} - 1
	\end{equation}
	and \( f_R \) and \( f''_R \) are calculated using the Hu-Sawicki model's form of \( f(R) \).
	
	\item \textbf{For \( \xi_4 \):}
	\begin{equation}
		\frac{d\xi_4}{dN} = \xi_4 - \xi_4 \xi_1
	\end{equation}
	where \( \xi_4 \) describes the evolution of perturbations in the scalar field.
	
	\item \textbf{For \( \xi_5 \):}
	\begin{equation}
		\frac{d\xi_5}{dN} = -\xi_5 - \xi_5 \xi_3
	\end{equation}
	where \( \xi_5 \) corresponds to the matter density contribution.
	
	\item \textbf{For \( \xi_6 \):}
	\begin{equation}
		\frac{d\xi_6}{dN} = \epsilon_2 - \xi_6 \xi_1 - \epsilon_1 \xi_6
	\end{equation}
	where \( \xi_6 \) describes the gravitational potential's evolution.
	
	\item \textbf{For \( \xi_7 \):}
	\begin{equation}
		\frac{d\xi_7}{dN} = \xi_6 - \xi_7 \xi_1
	\end{equation}
	where \( \xi_7 \) is related to the gravitational potential.
	
	\item \textbf{For \( \xi_8 \):}
	\begin{equation}
		\frac{d\xi_8}{dN} = \xi_8 (3 \omega_\nu - 1) - \xi_3 \xi_8 + \Gamma \xi_2 \xi_8
	\end{equation}
	where \( \xi_8 \) involves the neutrino contribution, and \( \Gamma \) is a coupling term.
\end{itemize}

Where  $N=lna$ thus,$\frac{d}{dN}= \frac{1}{H} \frac{d}{d\eta}$. 

The function $\beta $ is crucial in the study of modified gravitational theories, particularly in the context of $f(R)$ gravity. This parameter encapsulates the effects of modifications to General Relativity (GR) introduced by the function $f(R)$, which represents additional degrees of freedom in the gravitational action. The term $f''_R$ denotes the second derivative of the function $f(R)$ with respect to the Ricci scalar $R$, which plays a significant role in determining the dynamics of the scalar degree of freedom associated with $f(R)$ gravity. The  $\beta$ essentially quantifies the deviation from GR by comparing the contributions of the modified gravity term to the standard Hubble expansion rate $H$. A value of $\beta = -1$ would correspond to no modification, i.e., standard GR. The behavior of this function influences the cosmic evolution and the structure formation, affecting observables such as the growth rate of cosmic structures and the dynamics of dark energy. Studies such as those by \cite{DeFelice2010, Sotiriou2010, Nojiri2011} provide comprehensive reviews of these modified theories and their implications on cosmological phenomena.

After some calculation from equations, for simplicity, we can obtain the above parameters in terms of the new variables as 

\begin{equation}\label{is}
	\epsilon_{1}=\frac{\mathcal{H}^{\prime}}{\mathcal{H}^{2}}, \epsilon_{2}=\frac{\Psi^{\prime\prime}}{\phi H^{2}}, \epsilon_{3}=\frac{\phi^{\prime\prime}}{\phi H^{2}}, \epsilon_{4}=\frac{\delta^{\prime}}{\phi H}
\end{equation}

Now, for the autonomous equations of motions, we obtain

\begin{equation}\label{is}
	\begin{split}
		\epsilon_{1}=\frac{1}{1-\xi_{7}}[\xi_{1}+\xi_{6}+\frac{1}{3}\xi_{2}^{2}(1+\xi_{7})+(3-\xi_{7}+\xi_{6})\xi_{5}+\\(3-\xi_{7}+\xi_{6})\xi_{8}-\frac{\kappa^{2}}{k^{2}}\xi_{5}\xi_{4}\xi_{2}^{2}]
	\end{split}
\end{equation}

\begin{equation}\label{eq4}
	\begin{split}
		\epsilon_{2}=\frac{-2}{1-\xi_{7}}[\xi_{1}+\xi_{6}+\frac{1}{3}\xi_{2}^{2}(1+\xi_{7})+\\(3-\xi_{7}+\xi_{6})\xi_{5}- \frac{\kappa^{2}}{k^{2}}\xi_{5}\xi_{4}\xi_{2}^{2}] \\
		-\xi_{1}-3\xi_{6}+\frac{1}{3}\xi_{2}^{2}-\xi_{6}\xi_{1}+\\ \frac{1}{3}\xi_{2}^{2}(1-2\xi_{7})+\frac{\Omega}{3}(1-\xi_{7})\frac{1}{k^{2}}\xi_{2}^{2}
	\end{split}
\end{equation}

\begin{equation}\label{is}
	\begin{split}
		\epsilon_{3}=-\epsilon_{2}-3\epsilon_{1}(1+\frac{1}{3}\xi_{7})-3\xi_{1}-3\xi_{6}-2\xi_{7}-\\ (3-\xi_{7}+3\xi_{1})\xi_{5}+\beta(\xi_{7}-3)
	\end{split}
\end{equation}

\begin{equation}\label{is}
	\epsilon_{4}=\frac{-\kappa^{2}[(2-\xi_{7})\xi_{3}+\xi_{1}+\xi_{6}+1+\xi_{7}]}{\kappa^{2}\xi_{5}}+3\xi_{6}
\end{equation}
The parameter $\epsilon_{1}$ is of great importance, as it allows the expression of fundamental cosmological parameters such as the deceleration parameter \( q \) and the effective equation of state (\( w_{\text{eff}} \)) in terms of it. Specifically, \( q = -1 - \frac{\mathcal{H}'}{\mathcal{H}^2} \) and \( w_{\text{eff}} = -1 - \frac{2}{3}\frac{\mathcal{H}'}{\mathcal{H}^2} \).
The deceleration parameter is a dimensionless parameter that characterizes the rate at which the expansion of the Universe is slowing down. It is defined as the negative of the ratio of the cosmic acceleration to the cosmic expansion rate squared. Mathematically, it is expressed as:
$q =  - \frac{{a''a}}{{{{a'}^2}}}$
where \( a \) is the scale factor of the Universe, \( {a'} \) represents the first derivative of the scale factor with respect to cosmic time, and \( {a''} \) represents the second derivative.

\section*{Modifications to the CLASS Code for the Hu-Sawicki Model}

In this work, we have made significant modifications to the CLASS code to incorporate the Hu-Sawicki \( f(R) \) gravity model. These changes were necessary to account for the additional free parameters introduced by the model, particularly those related to the evolution of cosmological perturbations and the modified growth of structures. The key modifications are summarized as follows:

1. **Implementation of Autonomous Equations**: We integrated the autonomous system of differential equations governing the evolution of cosmological parameters into the CLASS code. These equations describe the dynamics of variables such as \( \xi_i \) and \( \beta \), where \( \beta \) characterizes the deviations from General Relativity, specifically due to the second derivative of the \( f(R) \) function.

2. **Modified Growth Parameters**: The modified growth parameters, expressed in terms of the \( \xi_i \) variables, were incorporated into CLASS to compute key observables like the growth rate of cosmic structures and expansion history. This allows for a more precise analysis of the effects of \( f(R) \) gravity on the evolution of the universe.

3. **Friedmann Equation Updates**: The Friedmann equations were updated to include the modification due to \( f(R) \), particularly the \( \beta \) function. This required adapting the CLASS framework to properly account for the modified Hubble parameter and ensure accurate cosmological predictions.

These modifications enable CLASS to simulate cosmological observables in the Hu-Sawicki \( f(R) \) framework, facilitating a robust analysis of the model's impact on the Hubble tension and other cosmological phenomena.

In the next section we investigate the model numerically.

\section{Numerical Analysis}

\subsection{Codes}

To evaluate the success of the model under study, we perform a series of Markov-chain Monte Carlo (MCMC) runs, using the public code { MontePython-v3}\footnote{\url{https://github.com/brinckmann/montepython_public}}\cite{54,55}, which we interface with our modified version of { CLASS}~\cite{56,57}. To test the convergence of the MCMC chains, we use Gelman-Rubin \cite{58} criterion $|R -1|\!\lesssim\!0.01$. To post-process the chains and plot figures we use {\sf GetDist} \cite{59}.  We pay particular attention to the impact of coupled neutrinos on the evolution of the Universe.

\section{Data Integration and Likelihood Functions in MontePython}
\subsection{Data}
The observational data where used in this study are:\\
$\bullet$ Pantheon catalog:
We used an  updated Pantheon + Analysis catalog consisting of 1701  SNe Ia covering the redshift range $0.001 < z < 2.3$\cite{42}.\\
$\bullet$ {CMB data}:
We used the latest large-scale MB temperature and
polarization angular power spectra from the final release of Planck 2018 plikTTTEEE+lowl+lowE
\cite{43}.\\ 
$\bullet$ {CC data}: The 32 $H(z)$ measurements listed in Table I have a redshift range of $0.07 \leq z \leq 1.965$ (\cite{zhang2014,borghi2022,ratsimbazafy2017,stern2009,Moresco3}). The covariance matrix of the 15 correlated measurements originally from Refs. (\cite{Moresco,Moresco1,Moresco2}) , discussed in Ref. \cite{Moresco3}, can be found at https://gitlab.com/mmoresco/CCcovariance/.\\
$\bullet$ {BAO data}:
We also used various measurements of BAO data  (\cite{Carter2018,Gil-Marin2020,Bautista2021,DES2022,Neveux2020,Hou2021,Bourboux2020}).\\
$\bullet$ { Lensing data}: we consider the 2018 CMB  Lensing reconstruction power spectrum data,
obtained with a CMB trispectrum analysis in \cite{Aghanim1}.

\subsection{Cosmic Microwave Background (CMB)}

MontePython utilizes CMB data to constrain cosmological models by comparing theoretical predictions of the CMB power spectra with observed power spectra. The likelihood function for CMB data, $\mathcal{L}_{\text{CMB}}$, is based on the deviation between the observed and theoretical power spectra. It is given by:
\begin{equation}
	\mathcal{L}_{\text{CMB}} \propto \exp\left(-\frac{1}{2} \left[ \textbf{d}_{\text{obs}} - \textbf{d}_{\text{theory}} \right]^T \textbf{C}^{-1} \left[ \textbf{d}_{\text{obs}} - \textbf{d}_{\text{theory}} \right] \right),
\end{equation}
where $\textbf{d}_{\text{obs}}$ and $\textbf{d}_{\text{theory}}$ represent the observed and theoretical CMB power spectra, respectively, and $\textbf{C}$ is the covariance matrix accounting for measurement uncertainties and correlations. MontePython incorporates precomputed CMB likelihood functions, typically provided by analysis pipelines such as those from the Planck mission.

\subsection{Cosmic Chronometers (CC)}

MontePython uses Cosmic Chronometers data, which provides measurements of the expansion rate $H(z)$ at various redshifts, to constrain model parameters. The likelihood function $\mathcal{L}_{\text{CC}}$ is expressed as:
\begin{equation}
	\mathcal{L}_{\text{CC}} \propto \exp\left(-\frac{1}{2} \sum_{i} \left[ \frac{H(z_i)_{\text{model}} - H(z_i)_{\text{data}}}{\sigma_{H}(z_i)} \right]^2 \right),
\end{equation}
where $H(z_i)_{\text{model}}$ denotes the predicted expansion rate at redshift $z_i$, $H(z_i)_{\text{data}}$ is the observed expansion rate, and $\sigma_{H}(z_i)$ is the uncertainty. MontePython integrates these measurements by specifying them in the configuration file and computing the likelihood based on model predictions.

\subsection{Baryon Acoustic Oscillations (BAO)}

MontePython incorporates BAO data by fitting theoretical predictions of the BAO scale with observed measurements. The likelihood function $\mathcal{L}_{\text{BAO}}$ is given by:
\begin{equation}
	\mathcal{L}_{\text{BAO}} \propto \exp\left(-\frac{1}{2} \frac{\left(D_V^{\text{model}} - D_V^{\text{data}}\right)^2}{\sigma_{D_V}^2}\right),
\end{equation}
where $D_V^{\text{model}}$ is the model’s predicted BAO scale, $D_V^{\text{data}}$ is the observed scale, and $\sigma_{D_V}$ is the uncertainty in the measurement. In MontePython, users input observed BAO scales and their uncertainties, and the software computes the likelihood based on the fit between theoretical predictions and observations.

\subsection{Pantheon Supernovae}

MontePython uses Pantheon supernovae data by comparing the theoretical distance moduli with observed moduli. The likelihood function $\mathcal{L}_{\text{SN}}$ for supernovae data is expressed as:
\begin{equation}
	\mathcal{L}_{\text{SN}} \propto \exp\left(-\frac{1}{2} \sum_{j} \left[ \frac{\mu_{j}^{\text{model}} - \mu_{j}^{\text{data}}}{\sigma_{\mu}(j)} \right]^2 \right),
\end{equation}
where $\mu_{j}^{\text{model}}$ is the predicted distance modulus for the $j$-th supernova, $\mu_{j}^{\text{data}}$ is the observed distance modulus, and $\sigma_{\mu}(j)$ is the uncertainty. MontePython calculates the likelihood for each parameter set by comparing the model's predicted distance moduli with the observed values.

\subsection{Lensing Data}

MontePython integrates lensing data by evaluating the theoretical predictions of the lensing power spectrum against observed data. The likelihood function for lensing data, $\mathcal{L}_{\text{lensing}}$, is expressed as:
\begin{align}
	\mathcal{L}_{\text{lensing}} &\propto \exp\left(-\frac{1}{2} \left[ \textbf{d}_{\text{lensing,obs}} - \textbf{d}_{\text{lensing,theory}} \right]^T \right. \nonumber \\
	&\quad \times \left. \textbf{C}_{\text{lensing}}^{-1} \left[ \textbf{d}_{\text{lensing,obs}} - \textbf{d}_{\text{lensing,theory}} \right] \right),
\end{align}

where $\textbf{d}_{\text{lensing,obs}}$ and $\textbf{d}_{\text{lensing,theory}}$ are the observed and theoretical lensing power spectra, respectively, and $\textbf{C}_{\text{lensing}}$ is the covariance matrix incorporating measurement uncertainties and correlations. This data is crucial for understanding the distribution of dark matter and the large-scale structure of the universe.

\subsection{MCMC Sampling with MontePython}

In MontePython, the likelihood functions for all data types are combined to form the total likelihood function:
\begin{equation}
	\mathcal{L}(\text{data} | \theta) = \mathcal{L}_{\text{CMB}} \times \mathcal{L}_{\text{CC}} \times \mathcal{L}_{\text{BAO}} \times \mathcal{L}_{\text{SN}} \times \mathcal{L}_{\text{lensing}},
\end{equation}
where $\theta$ represents the cosmological parameters.

MontePython performs Markov Chain Monte Carlo (MCMC) sampling to explore the parameter space. The posterior probability distribution is given by:
\begin{equation}
	\mathcal{P}(\theta | \text{data}) \propto \mathcal{L}(\text{data} | \theta) \cdot \text{Prior}(\theta),
\end{equation}
where $\text{Prior}(\theta)$ is the prior distribution for the parameters. MontePython iterates over parameter values, proposing new sets of parameters, and accepting or rejecting them based on their likelihood. This process generates a chain of parameter samples, from which posterior distributions and constraints on the parameters are derived.

 \section{constraint on total mass of neutrinos}

The behavior of neutrinos and elusive and enigmatic particles are significantly influenced by the modified gravitational theories encapsulated in perturbed $f(R)$ gravity \cite{Nojiri2011}. In this section we study constraint on the total mass of neutrinos. The equation relates the energy density of neutrinos (\(\rho_{\nu}\)), the scale factor (\(a\)), and the modification term \((1+f_{R})\)is given by \cite{DeFelice2010,Sotiriou2010}:
  \begin{equation}
  	\xi_{9} = \frac{\rho_{\nu}a^{2}}{(1+f_{R})}.
  \end{equation}
  
  To estimate the mass of neutrinos using this equation, additional information or assumptions are necessary. The energy density of neutrinos is expressed in terms of their mass (\(m_{\nu}\)) and temperature (\(T_{\nu}\)). A common approach involves using the Fermi-Dirac distribution for relativistic neutrinos \cite{Kolb1990,Lesgourgues2006}:
  \begin{equation}
  	\rho_{\nu} = \frac{7\pi^2}{120} g_{\nu} T_{\nu}^4
  \end{equation}
where \(g_{\nu}\) denotes the number of degrees of freedom for neutrinos, with a value of 2 for each neutrino species. 
  Substituting this expression for \(\rho_{\nu}\) into the original equation, you obtain:
  \begin{equation}
  	\xi_{9} = \frac{\left(\frac{7\pi^2}{120} g_{\nu} T_{\nu}^4\right) a^{2}}{(1+f_{R})}
  \end{equation}
  
  To solve for the neutrino mass (\(m_{\nu}\)), specific values for parameters and potentially additional assumptions are needed. It's important to note that cosmological models can vary, and the approach may depend on the assumptions made in the model. The relationship between $\Omega_{\nu}$ and the sum of neutrino masses is given by \cite{Lesgourgues2006}:
  \begin{equation}
  	{\Omega _\nu } = \frac{{{{\sum m }_\nu }}}{{94{h^2}}}
  \end{equation}

To determine the total mass of neutrinos, ${\sum m }_\nu $, we need to employ MCMC method to find the best-fitting values for the cosmological parameters $\Omega_{\nu}$ and $h$. The elusive nature of neutrinos makes their direct measurement challenging. Assigning prior distributions to parameters, and iteratively sampling the parameter space, the MCMC approach entails defining a cosmological model with relevant parameters, constructing a likelihood function based on observed data. The best-fitting values of $\Omega_{\nu}$ and $h$ are determined by identifying the region where the likelihood is maximized. These optimal values are then used in the cosmological model to calculate the elusive total mass of neutrinos, shedding light on their contribution to the cosmic structure.

The numerical values of $ \sum m_{\nu}$ with $95$\% $CL$ for different data combination is given in Table II that with the combination of all dataset we have $\sum m_{\nu}< 0.121 eV$
\begin{table*}[ht!] 
	\caption{Numerical Values for \(z_{nr}\), \(z_{DA}\), \(\Gamma\), and \(\sum m_{\nu}\)}
	\resizebox{0.9\textwidth}{!}{
		\begin{tabular}{|c|c|c|c|c|c|} 
			\hline
			\textbf{Data Set} & \textbf{\(z_{nr}^{\text{NH}}\)} & \textbf{\(z_{nr}^{\text{IH}}\)} & \textbf{\(z_{DA}\)} & \textbf{\(\Gamma\)} & \textbf{\(\sum m_{\nu}\)} \\ 
			\hline
			CMB + Lensing & 259.47 & 719.7 & \(0.62 \pm 0.022\) & \(0.61 \pm 0.27\) & \(< 0.38\) \\ 
			\hline
			CMB + BAO + Lensing & 154.46 & 310.12 & \(0.51 \pm 0.018\) & \(0.641 \pm 0.055\) & \(< 0.164\) \\ 
			\hline
			CMB + CC + Lensing & 189.44 & 380.68 & \(0.48 \pm 0.026\) & \(0.642 \pm 0.062\) & \(< 0.201\) \\ 
			\hline
			CMB + Pantheon + Lensing & 264.77 & 430.30 & \(0.49 \pm 0.03\) & \(0.62 \pm 0.21\) & \(< 0.28\) \\ 
			\hline
			CMB + CC + BAO + Pantheon + Lensing & 113.66 & 229.30 & \(0.49 \pm 0.024\) & \(0.639 \pm 0.054\) & \(< 0.121\) \\ 
			\hline
	\end{tabular}}
\end{table*}

In Fig. 1, the constraints on $\sum m_{\nu}$, and  $H_0$ from the variety of dataset combination in perturbed $f(R)+m_{\nu}$ model are shown. With respect to the minimum sum of neutrino mass from the observations of neutrino oscillations,$\sum m_{\nu}= 0.06 eV$, one can see, that $H_0$ constraints strongly depends on  $\sum m_{\nu}$ in all data combination as long as CMB data is considered. 

\begin{figure}
	\includegraphics[width=8.5 cm]{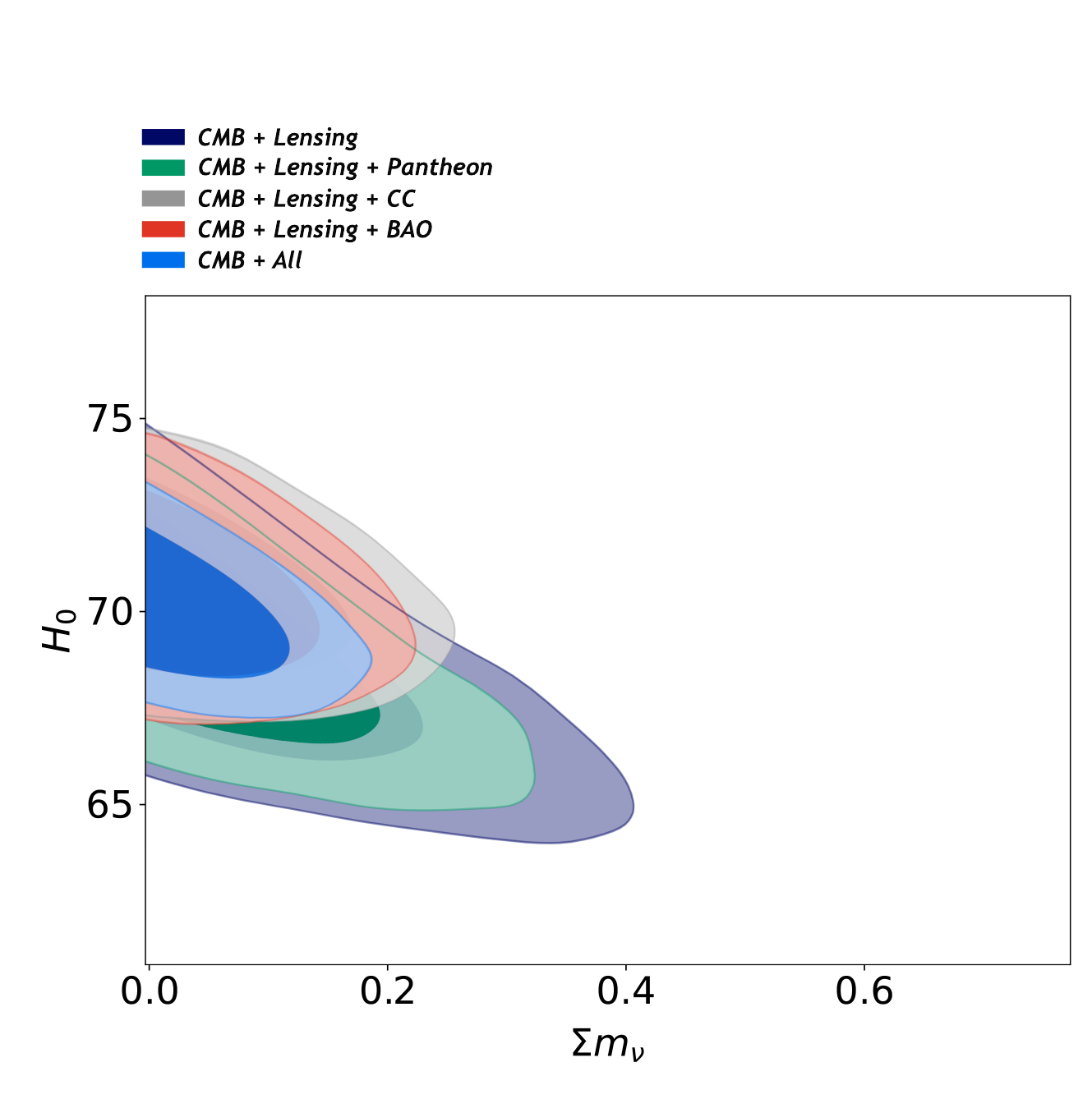}
	\vspace{-0.12cm}
	\caption{\small{The constraints at the (95$\% $CL.) two-dimensional contours for $\sum m_{\nu}$ and $H_0$ in perturbed $f(R)+m_{\nu}$for different combination of data }}\label{fig:omegam2}
\end{figure}

Similarly, the analysis, incorporating various combination of cosmological datasets, reveals distinct values for \(\Gamma\) and shown in Table I. Noting that, the results derived from the specific form of the external source term  in equation (7) where \(\Gamma\) plays a crucial role in characterizing the interaction between $f(R)$ gravity and neutrinos. The constraints on model interaction parameter ($\Gamma$), and Hubble constant $H_0$ with 95\% CL from the variety of dataset combination are also shown in Fig. 2. 

 \begin{figure}
   	\includegraphics[width=8.5 cm]{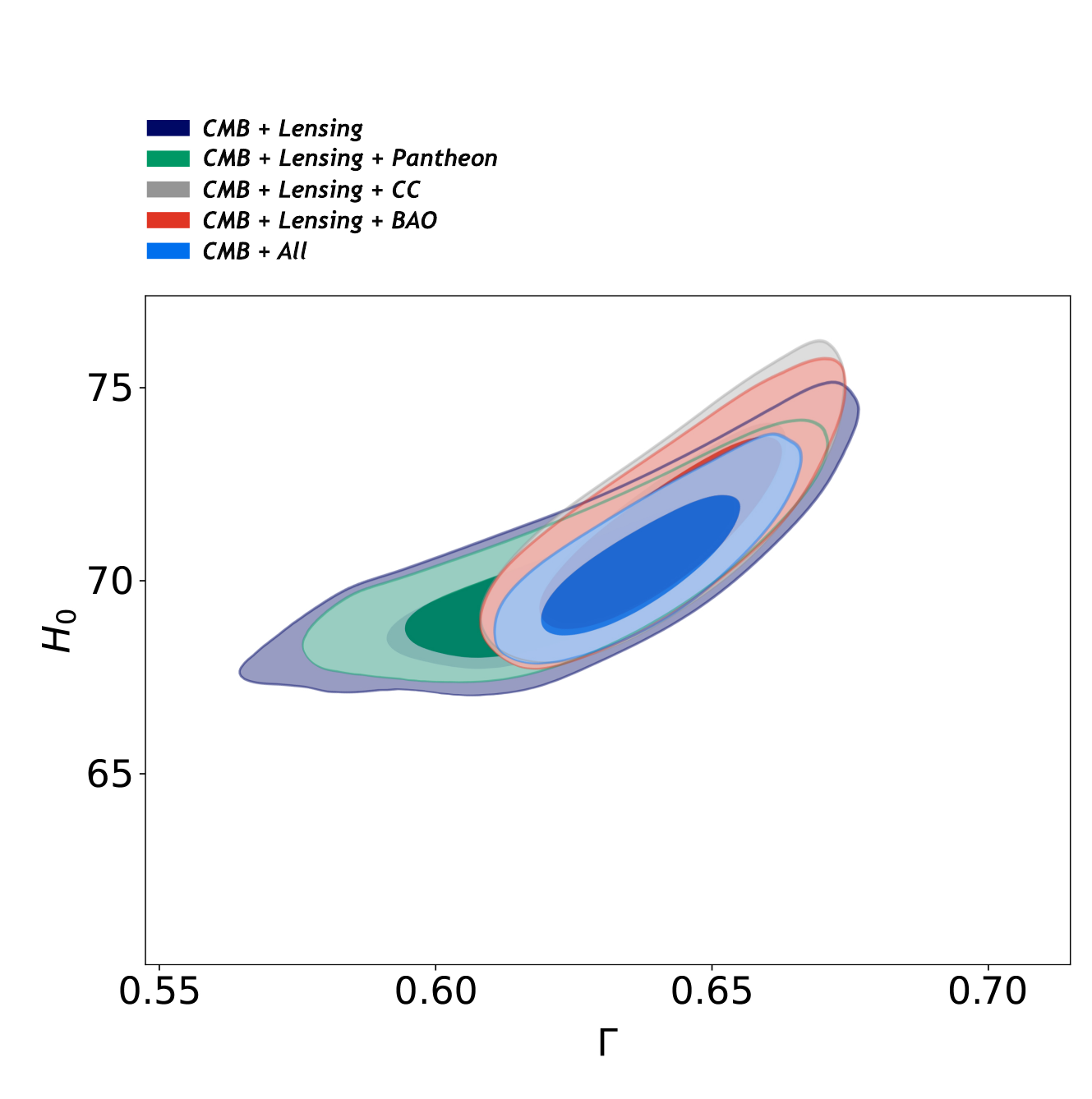}
   	\vspace{-0.12cm}
   	\caption{\small{The constraints at the (95$\% $CL.) two-dimensional contours for interaction parameter ($\Gamma$), and $H_0$ in perturbed $f(R)+m_{\nu}$ for different combination of data }}\label{fig:omegam2}
   \end{figure}    

{\bf The Relativistic to Non - Relativistic Neutrino Phase Transition, $z_{nr}$}

Estimating the relativistic to non-relativistic neutrino phase transition involves understanding the transition from the early Universe, where neutrinos are highly relativistic, to later epochs where they become non-relativistic. The transition is associated with the decoupling of neutrinos from the thermal bath of other particles in the primordial plasma. To estimate this transition, we can use the concept of freeze-out \cite{Dolgov2002}.

In the early Universe, neutrinos are in thermal equilibrium with other particles, maintained by frequent scattering interactions. As the Universe expands and cools, the temperature decreases, and the interaction rate drops. At a certain temperature, neutrinos decouple, and their distribution function freezes out \cite{Kolb1990, Dolgov2002}.

In the following, we estimate the  neutrinos transition from relativistic to non-relativistic at
redshift $z_{nr}$. Neutrinos decouple from the primordial plasma in a Fermi-Dirac distribution \cite{Lesgourgues2006}:
\begin{equation}
	f({p_\nu },{T_\nu }) = [\exp (\frac{{{p_\nu }}}{{{T_\nu }}}) + 1]^{-1}
\end{equation}
with temperature $T_\nu$ and average momentum 
$\left\langle {{p_\nu }} \right\rangle  = 3.15{T_\nu }$. One can express \(T_{\nu}\) in terms of CMB temperature and redshift by:
\begin{equation}
	{T_\nu }^0 = {(\frac{4}{{11}})^{\frac{1}{3}}}{T^0}_{CMB}
\end{equation}
where CMB temperature is $2.725K$. Massive neutrinos become non-relativistic when $p_\nu$ falls below their rest mass. 

Using a CMB temperature of $2.725K$ and given that in general $ T(z) = {T_0}(1 + z) $, we can then estimate the redshift at which a neutrino of mass  will become non-relativistic as:
\begin{equation}
	{z_{nr}} = (\frac{{{m_\nu }}}{{5.28 \times {{10}^{ - 4}}ev}}) - 1
\end{equation} 
The results obtain for  ${z_{\rm nr}}$ in different combination of dataset given in Table I.
The results obtained for ${z_{\rm nr}}$ using different combinations of datasets are summarized in Table I. When considering the neutrino mass structure for both the normal hierarchy (NH) and inverted hierarchy (IH), distinct distributions are applied \cite{Yarahmadi}. In the NH, the third neutrino mass is zero, while in the IH, the first and second masses are nearly zero. Specifically, in the NH, the total neutrino mass is divided between the remaining two masses (\(m_3 = 0\)), whereas in the IH, all the neutrino mass is attributed to the third neutrino (\(m_1 = m_2 = 0\)).

Using the NH configuration, the neutrino masses are \(m_{\nu}^{\text{NH}} = \sum m_{\nu}/2\), while for the IH configuration, the neutrino mass is \(m_{\nu}^{\text{IH}} = \sum m_{\nu}\). Consequently, the calculated \(z_{\rm nr}\) values vary across datasets. For CMB + Lensing, \(m_{\nu}^{\text{NH}} = 0.19 \, \text{eV}\) gives \(z_{\rm nr}^{\text{NH}} = 259.47\), while \(m_{\nu}^{\text{IH}} = 0.38 \, \text{eV}\) results in \(z_{\rm nr}^{\text{IH}} = 719.7\). For CMB + BAO + Lensing, \(m_{\nu}^{\text{NH}} = 0.082 \, \text{eV}\) gives \(z_{\rm nr}^{\text{NH}} = 154.46\), and \(m_{\nu}^{\text{IH}} = 0.164 \, \text{eV}\) leads to \(z_{\rm nr}^{\text{IH}} = 310.12\).

Further analysis with CMB + CC + Lensing provides \(m_{\nu}^{\text{NH}} = 0.1005 \, \text{eV}\), yielding \(z_{\rm nr}^{\text{NH}} = 189.44\), while \(m_{\nu}^{\text{IH}} = 0.201 \, \text{eV}\) gives \(z_{\rm nr}^{\text{IH}} = 380.68\). For CMB + Pantheon + Lensing, \(m_{\nu}^{\text{NH}} = 0.14 \, \text{eV}\) results in \(z_{\rm nr}^{\text{NH}} = 264.77\), and \(m_{\nu}^{\text{IH}} = 0.28 \, \text{eV}\) gives \(z_{\rm nr}^{\text{IH}} = 430.30\). These results highlight the impact of dataset combinations on \(z_{\rm nr}\) estimates under different neutrino hierarchies.

Finally, for the comprehensive combination of CMB + CC + BAO + Pantheon + Lensing, \(m_{\nu}^{\text{NH}} = 0.0605 \, \text{eV}\) gives \(z_{\rm nr}^{\text{NH}} = 113.66\), and \(m_{\nu}^{\text{IH}} = 0.121 \, \text{eV}\) results in \(z_{\rm nr}^{\text{IH}} = 229.30\). These recalculations demonstrate the sensitivity of \(z_{\rm nr}\) to both the assumed hierarchy of neutrino masses and the observational datasets employed, emphasizing the importance of careful dataset selection and hierarchy consideration in cosmological analyses.

{\bf Deceleration to Acceleration Phase Transition, $z_{\rm DA}$}

The coupling of neutrinos with \(f(R)\) gravity creates a compelling approach to understanding cosmic evolution, especially during the pivotal transition from deceleration to acceleration. This transition is crucial, as it signifies the Universe's shift from being dominated by matter to being driven by dark energy, a shift initially observed in studies of distant supernovae.

In standard cosmology, the Universe expands at varying rates: it decelerates during the radiation and matter-dominated eras but accelerates as dark energy becomes more influential. The modified Friedmann equations, influenced by the coupling, alter the Universe's expansion trajectory.

This coupling affects neutrinos' effective energy density and pressure, impacting the overall energy budget of the Universe and, consequently, its rate of expansion. The phase transition from deceleration to acceleration within this framework depends on the specifics of the \(f(R)\) gravity and the nature of the neutrino interaction. The transition's timing and dynamics are influenced by these additional degrees of freedom.
To derive the redshift \( z_{\rm DA} \) at which the transition from deceleration to acceleration occurs in the context of perturbed \( f(R) \) gravity, we start with the deceleration parameter \( q = -1 - \epsilon_1 \) and set \( q = 0 \), leading to \( \epsilon_1 = -1 \). Expressing \( \epsilon_1 \) in terms of normalized parameters \( \xi_i \), we have \( \epsilon_1 \). Setting this equal to -1 yields a rearranged equation: \( - (1 - \xi_7) = \xi_1 + \xi_6 + \frac{1}{3} \xi_2^2 (1 + \xi_7) + (3 - \xi_7 + \xi_6) \xi_5 + (3 - \xi_7 + \xi_6) \xi_8 - \frac{\kappa^2}{k^2} \xi_5 \xi_4 \xi_2^2 \). The relationship between the scale factor \( a \) and redshift \( z \) is given by \( 1 + z = \frac{1}{a} \), which implies at \( z_{\rm DA} \), \( a = \frac{1}{1 + z_{\rm DA}} \). By substituting the appropriate expressions for \( \xi_i \) into this modified equation and solving, we can express \( z_{\rm DA} \) as a function of these parameters, providing insight into the transition dynamics in the modified gravity framework.

\begin{figure*}
	\includegraphics[width=15.5 cm]{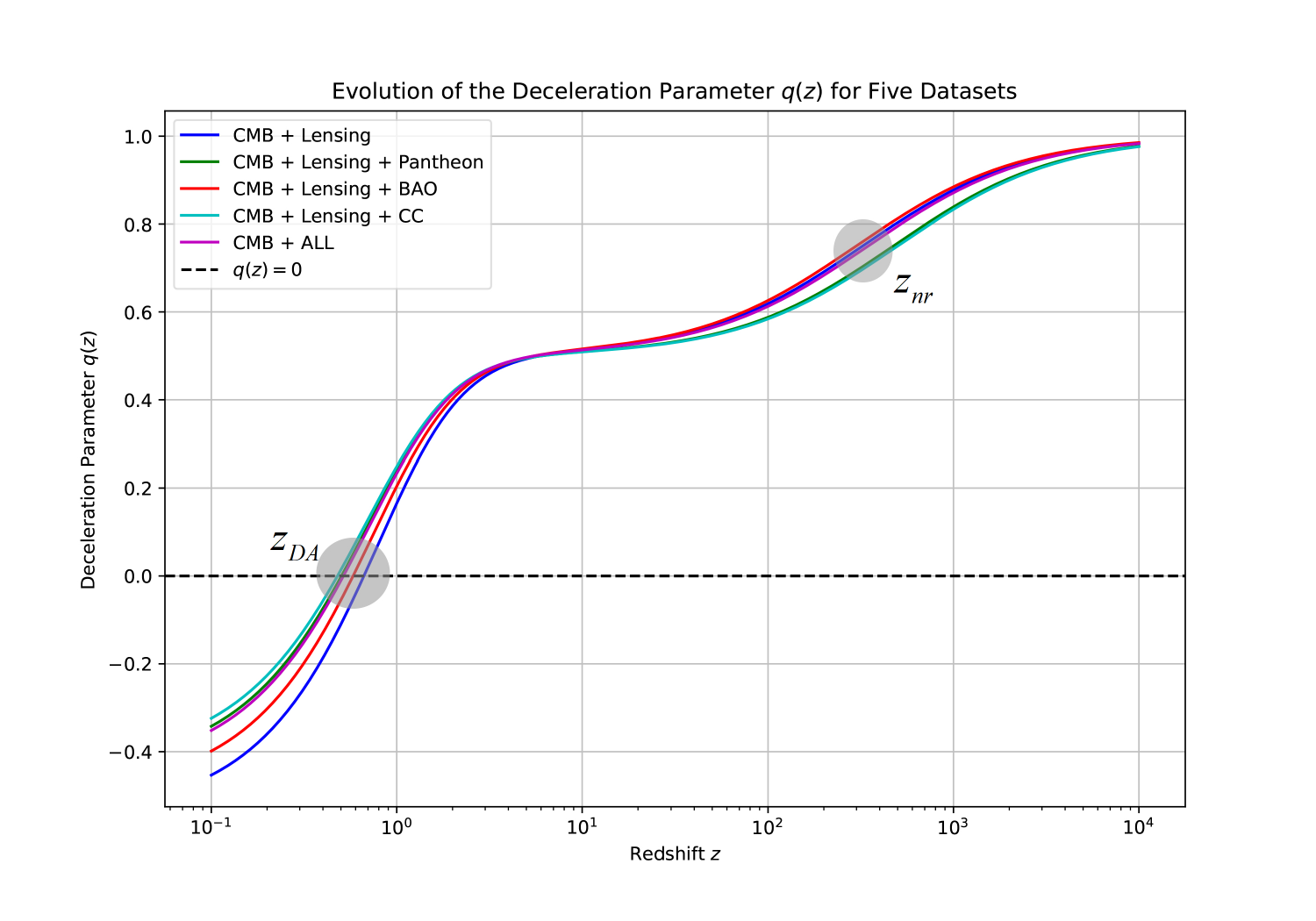}
	\vspace{-0.02cm}
	\caption{\small{The evolution of deceleration parameter q with respect to redshift z for different combination of dataset. }}\label{fig:omegam2}
\end{figure*}

The results obtain for Deceleration - Acceleration phase transition redshift ${z_{\rm DA}}$ in different combination of dataset are also given in Table I and shown in Fig. 4. The results are in  good agreement with those obtained in   \cite{dos Santos}  and \cite{53}, where the value of 
$z_{\rm DA}=0.46 $ was calculated. 

  \section{Structure Formation of The  Early Universe}
  The coupling of neutrinos with $f(R)$ gravity can have notable effects on CMB power spectrum.
  The study involves a detailed examination of the altered field equations, modified Friedmann equations, and perturbation equations governing both neutrinos and the metric within the context of $f(R)$ gravity.
  
  The shifts in the CMB power spectrum peaks are attributed to changes in gravitational potentials, expansion history, and the growth of large-scale structures induced by the coupling. Figure 3 shows the effect of coupling of neutrinos with $f(R)$ gravity on CMB power spectrum. The inclusion of neutrino density impacts the overall densities of baryonic matter and cold dark matter, consequently altering the peaks in CMB in comparison to the standard 
  $\Lambda$CDM model. Specifically, variations in neutrino density induce shifts in the peaks of the CMB power spectrum due to their gravitational effects on the evolving structure of the universe. This phenomenon highlights the interplay between neutrino and cosmological observations, emphasizing the need for precise modeling and observational constraints in modern cosmology.
	
 The Sachs-Wolfe effect arises from the interaction between gravitational potentials and photons navigating evolving gravitational fields in an expanding Universe. The Neutrino coupling induces alterations in gravitational fields, leaving a distinctive imprint on CMB power spectrum. In Figure 3, we illustrate the Sachs-Wolfe effect for angular multi poles ($l$) ranging from 10 to 100. The comparative analysis includes scenarios with different data sets alongside the standard $\Lambda$CDM model. The figure illustrates that, for $l \geq 200$, the first peak experiences a downward shift compared to the $\Lambda$CDM model. The shifts in CMB power spectrum suggests potential impact on late-time acceleration of the Universe. This is noteworthy for smaller angular scales, where changes in the late-time acceleration induced by the model can manifest. The model introduces a transformative element to the behavior of acoustic oscillations in the primordial plasma. It results in shifts in the positions and amplitudes of peaks and troughs in the CMB power spectrum associated with acoustic oscillations. The observed downward shift of the first peak for $l \geq 200$ in Figure 3 underscores this influence \cite{HuSugiyama1995, Komatsu2011,EisensteinHu1998}.
  	
Furthermore, models contribute to a noticeable change in the distribution of matter in the Universe. The impact extends to the growth of large-scale structures, ultimately influencing the distribution of matter at the time of recombination and consequently shaping the CMB temperature fluctuations. The changes in the growth rate of structure are reflected in the higher multipoles of the CMB power spectrum, providing valuable insights into the intricate interaction between cosmic components and their effects on the observable Universe.

\begin{figure*}
  	\includegraphics[width=20 cm]{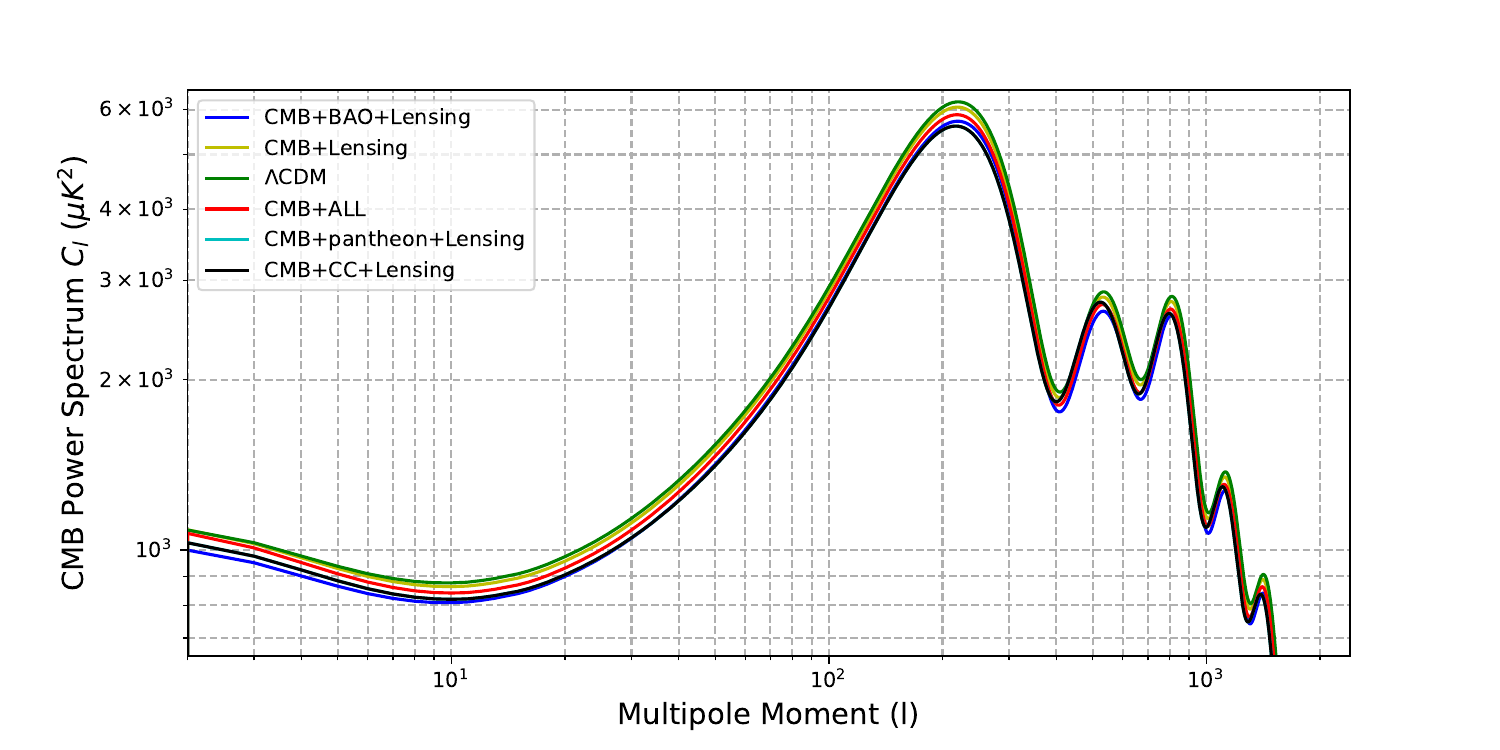}
  	\vspace{-0.12cm}
  	\caption{\small{Comparison the CMB power spectrum  between different combination of data sets  for perturbed $f(R)+m_\nu$ and  $\Lambda $CDM model. }}\label{fig:omegam2}
  \end{figure*}
  
In the following, we study the the model in structure formation on Early Universe by using the characteristic wavenumber \cite{Suarez}:

\begin{equation}
	k_J = \left(\frac{4\pi G\rho}{c_s^2}\right)^{1/2},
	\label{intro2}
\end{equation}

 where $c_s^2 = P'(\rho)$ represents the square of the speed of sound in the medium and the equation is referred to as the Jeans wavenumber. The Jeans length $\lambda_J = 2\pi/k_J$ offers an estimate of the minimum size of objects that can undergo gravitational collapse \cite{Suarez}. We constrain $c_s$ using MCMC method. The results for different combinations of data sets for $f(R)$ gravity are given in Table I. In matter dominated era $c_{s}^2 = 0$. These  results shows that the model can good explain the circumstance of structure formation in the Early Universe. Our results are in good agreement with \cite{Suarez}.

The Jeans Wavenumbers in our model for different combination of observational data are also given in Table III. These results are in general agreement with (\cite{Masatoshi};\cite{Christos}; \cite{Yarahmadi}).\\

\begin{table*}
	\caption{Numerical Values for $c_{s}$ and  $k_{J}$ }
	\resizebox{0.8\textwidth}{!}{
		\begin{tabular}{|c|c|c|c|c|}
			\hline
			\textbf{Data Set} & \textbf{sound speed ($c_{s})$ } &  \textbf{ Jeans wavenumbers($K_{J}) Mpc^{-1}c$}   \\
	    	\hline
		    CMB + Lensing &  $0.29 \pm 0.047$ & $0.000221\pm0.00018$   \\
			\hline
			CMB + BAO + Lensing &  $0.36 \pm 0.057$ & $0.000261\pm0.00013$   \\
			\hline
			CMB + CC + Lensing &   $0.4 \pm 0.048$ & $0.00027\pm0.00017$  \\
			\hline
			CMB + Pantheon + Lensing &  $0.33 \pm 0.047$& $0.000259\pm0.00016$  \\
			\hline
			CMB + CC + BAO + Pantheon + Lensing  &    $0.25 \pm 0.54$ & $0.000215\pm0.00011$  \\
		
			\hline
	\end{tabular}}
\end{table*}

\section{Hubble tension}

A mismatch between observed and predicted values of the Hubble constant has led to a rethinking of the standard cosmological model. The neutrino coupling in our model can offer a more comprehensive understanding of the cosmic expansion dynamics. Direct measurements of \(H_0\) are crucial, relying on observations and calculations grounded in the local Universe. This involves studying astrophysical objects and phenomena in our cosmic neighborhood to determine the current rate of expansion. Two primary methods for direct measurements are employed: (1) observing Cepheid variable stars in nearby galaxies, allowing the calculation of distances and recession velocities, and (2) using Type Ia supernovae as standardizable candles to estimate distances and recession velocities in distant galaxies. Additionally, data from CC and BAO contribute to Hubble parameter estimation. While these direct measurements provide insights into the local Universe, a tension arises when comparing their results with the Hubble constant derived from early Universe observations like CMB or large-scale structure. This suggests a potential discrepancy, motivating investigations into alternative gravitational frameworks such as \(f(R)\) gravity to refine our understanding and potentially reconcile the observed discrepancies. To investigate the Hubble Tension in late Universe for Pantheon data, we use 896 SNe in redshift interval $0.1<z<2.3$. This is because the Hubble tension is not inherently a discrepancy at redshift \(z=0\), but rather at \(z \approx 0.1\). For the local universe the comparison of $ H_{0}$ for different data is given in Table IV. As can be seen for all data combination we found that $ H_{0} = 70.29 \pm 1.01 $  $kms^{-1} Mpc^{-1}$ at 68\% CL, which is close to Planck 2018 results with a 2.49$ \sigma $ and 1.83$ \sigma $ with R22.

Figure 5 demonstrate the comparison of \(H_{0}\) measurements for different combinations of datasets with results of Planck 2018 and R22. The horizontal dashed lines represent the respective results from Planck 2018 and R22, providing benchmarks for contextual interpretation.

In addition, to obtaining a more accurate results, we recalculate the Hubble tension by including CMB data to the original as can be seen in Table IV.
 \begin{sidewaystable}[h]
 	\centering
 	\caption{Hubble Constant Measurements and Tension with Planck 2018 and R22 in perturbed $f(R)$ gravity coupled with neutrinos}
 	\begin{tabular}{|l|c|c|c|c|}
 		\hline
 		\textbf{Dataset} & \textbf{Hubble Constant} (\(H_0\) in \(km s^{-1} Mpc^{-1}\)) & \textbf{Tension with R22} (\( \sigma \)) & \textbf{Tension with Planck 2018} (\( \sigma \)) \\
 		\hline
 		CMB + Lensing            & \(69.48 \pm 2.3\)  & 1.41 & 0.88 \\
 		\hline
 		CMB + BAO + Lensing       & \(70.18 \pm 1.63\) & 1.48 & 1.63 \\
 		\hline
 		CMB + CC + Lensing        & \(70.96 \pm 2.2\)  & 0.85 & 1.58 \\
 		\hline
 		CMB + Pantheon + Lensing  & \(71.13 \pm 2.4\)  & 0.73 & 1.52 \\
 		\hline
 		CMB + ALL                 & \(70.42 \pm 2.25\) & 1.06 & 1.31 \\
 		\hline
 	\end{tabular}
 	\label{tab:hubble_tension_final}
 \end{sidewaystable}

   \begin{figure*}
	\includegraphics[width=15 cm]{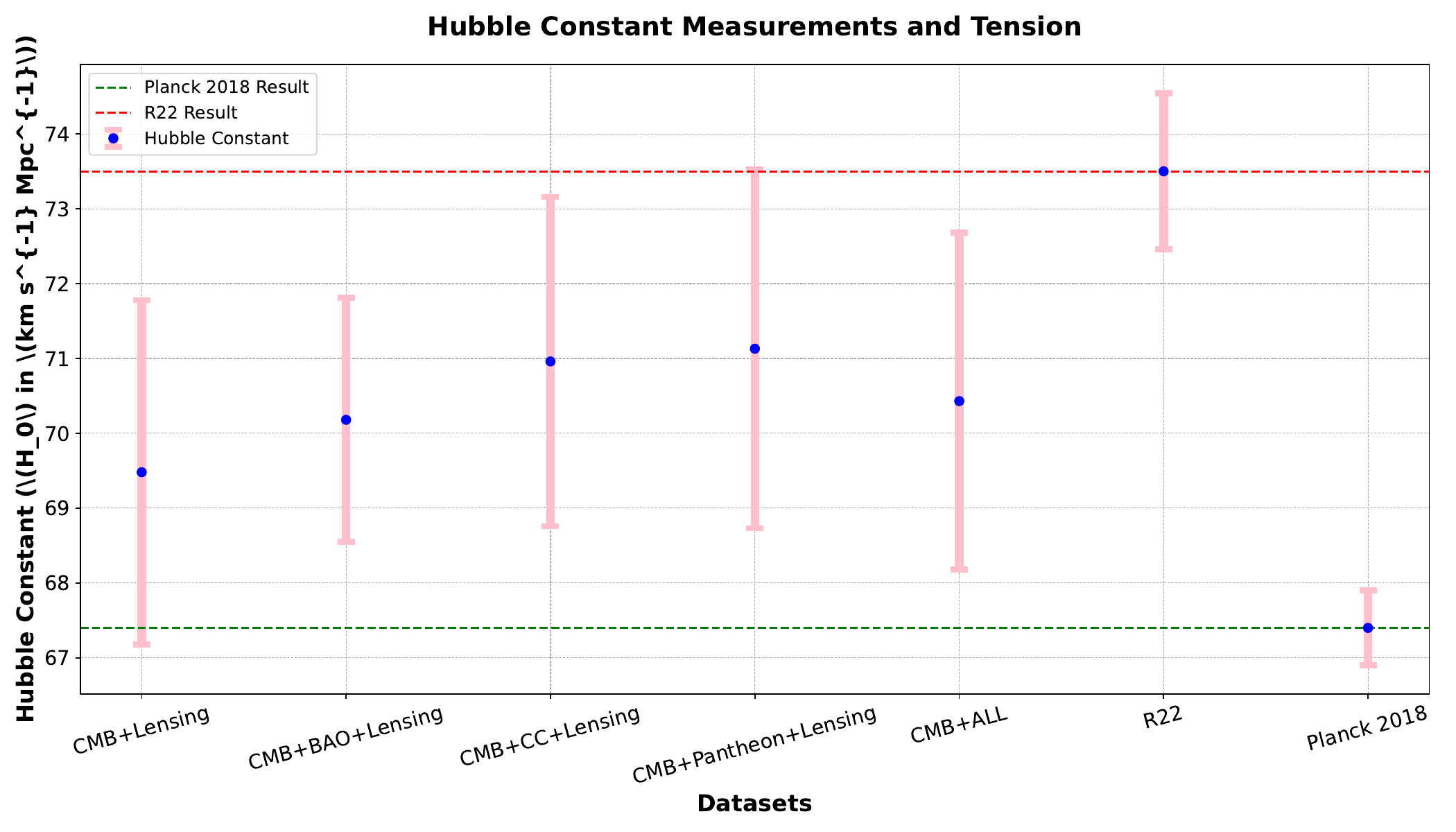}
	\vspace{-0.02cm}
	\caption{\small{The comparison of $H_{0}$ measurement for different combination of data sets(CMB+Other) with results of Planck 2018 and R22.
	}}\label{fig:omegam2}
\end{figure*}	
Comparing with $H_{0}$ is Table  III shows that adding Lensing data, alleviates $H_{0}$, these results are very close to \cite{Yarahmadi10,Ratra1}.

 \section{Summary and discution}

In this paper, we study interaction between perturbed $f(R)$ gravity and neutrinos, exploring 
 combination of CMB, CC, BAO, and Pantheon supernova datadatasets to constrain cosmological and model  parameters such as the  sum of neutrino masses ($\sum m_{\nu}$), sound speed ($c_s$), Jeans wavenumbers ($k_J$), relativistic to non - relativistic Neutrino Phase Transition ($z_{\rm nr}$), decelerating to accelerating cosmic expansion($z_{\rm DA}$) and interaction strength parameter, ($\Gamma$).  We also address the Hubble tension problem by measuring $H_0$ using different combination of dataset. The inclusion of neutrinos adds a layer of complexity to the analysis, influencing the strength of the interaction term ($\Gamma$) and, consequently, shaping the evolution of the cosmos. The analysis of the sum of neutrino masses ($\sum m_{\nu}$) from different dataset combinations reveals intriguing constraints. The results align broadly with those reported in \cite{43}. The sound speed ($c_s$) and Jeans wavenumbers ($k_J$) in the model exhibit intriguing behavior across different datasets. The transition from a decelerating to an accelerating cosmic expansion, characterized by the redshift $z_{\rm DA}$, is influenced by the intricate interplay between neutrinos, $f(R)$ gravity, and other cosmic components. Regarding the Hubble tension, using the CMB + Lensing + Pantheon + BAO + CC dataset, $H_0$ values at 68\% CL were obtained. These results, detailed in Table  III and illustrated in Figure 5, offer a comprehensive comparison of $H_0$ measurements for different dataset combinations with the results of Planck 2018 and R22. The \( f(R) \) gravity-neutrino coupling model presents several key advantages over traditional cosmological models, particularly the \(\Lambda\)CDM framework, making it a compelling alternative for understanding the universe's evolution and structure. This model's flexibility, due to its modification of General Relativity and inclusion of neutrino coupling, allows for a more dynamic explanation of cosmic acceleration and better captures deviations observed in large-scale structure formation. Notably, it provides tighter constraints on the sum of neutrino masses, offers a quantifiable interaction strength parameter (\(\Gamma\)), and enhances our understanding of the behavior of sound speed and Jeans wavenumbers, which are crucial for studying the stability and evolution of early universe perturbations. Additionally, it provides distinct and accurate redshift predictions for key cosmic epochs and has the potential to resolve the Hubble tension, offering \( H_0 \) values consistent with both local measurements and Planck data. Compared to other modified gravity models, the inclusion of neutrino coupling in our \( f(R) \) framework introduces an additional layer of interaction that enriches our understanding of cosmic dynamics. Thus, our model aligns well with current observational data, offers a more comprehensive understanding of the universe's fundamental properties, and paves the way for future cosmological research and discoveries, positioning itself as a superior and more predictive framework compared to \(\Lambda\)CDM and other existing models.

 \section*{Acknowledgments}
 The work of KB was partially supported by the JSPS KAKENHI Grant
 Number 21K03547 and 23KF0008.

\end{document}